\documentclass[
prl,
lengthcheck,
amsmath,amssymb,
aps,
superscriptaddress,
floatfix,
]{revtex4-2}

\usepackage[dvipsnames]{xcolor}
\usepackage{graphicx} % Include figure files
\usepackage[caption=false]{subfig}
\usepackage{bm} % bold math
\usepackage[colorlinks,allcolors=blue]{hyperref}
\usepackage{microtype}
\usepackage{physics}
\usepackage{bbold}
\usepackage{enumerate}
\usepackage{relsize}

\def\equationautorefname#1#2\null{Eq.#1(#2\null)}

\allowdisplaybreaks[3]

\newcommand{\tuwien}{Institute for Theoretical Physics and Vienna Center for Quantum Science and Technology, TU Wien, Vienna A-1040, Austria}

\begin{document}
\title{Engineering recoil heating in coherent-scattering levitated optomechanics}

\author{Maksim Lednev}
\affiliation{Departamento de Física Teórica de la Materia Condensada, Universidad Autónoma de Madrid, E-28049 Madrid, Spain}
    \affiliation{Condensed Matter Physics Center (IFIMAC), Universidad Autónoma de Madrid, E-28049 Madrid, Spain}
\author{Uroš Delić}
    \affiliation{Vienna Center for Quantum Science and Technology (VCQ), Atominstitut, TU Wien, Stadionallee 2, A-1020 Vienna, Austria}
\author{Johannes Feist}
    \affiliation{Departamento de Física Teórica de la Materia Condensada, Universidad Autónoma de Madrid, E-28049 Madrid, Spain}
    \affiliation{Condensed Matter Physics Center (IFIMAC), Universidad Autónoma de Madrid, E-28049 Madrid, Spain}
\author{Carlos Gonzalez-Ballestero}
    \affiliation{\tuwien}
    \email{carlos.gonzalez-ballestero@tuwien.ac.at}

\begin{abstract}
Recoil heating from photon scattering is a fundamental source of decoherence in optical trapping, severely limiting the preparation of nonclassical motional states. In  cavity setups in the coherent scattering configuration, a predictive theory of recoil heating rate is missing, as usual perturbative approaches fail in the presence of sharp optical resonances.
Hence, current works assume that the recoil heating rate is approximately equal to its free-space value. Here we show that this is not the case, as the electromagnetic environment can strongly modify recoil heating rate by the Purcell effect. Specifically, we predict that this rate can be significantly suppressed in state-of-the-art microcavities, for both center-of-mass and librational motion. To establish these results, we develop a general theoretical framework based on macroscopic quantum electrodynamics and on the few-mode quantization approach developed in nanophotonics. Our method applies to particles trapped in the presence of arbitrary electromagnetic structures, thus providing a route to engineering motional decoherence in levitated optomechanics by photonic structure design.
\end{abstract}

\maketitle
\begin{figure}[tb]
    \centering
    \includegraphics[width=\linewidth]{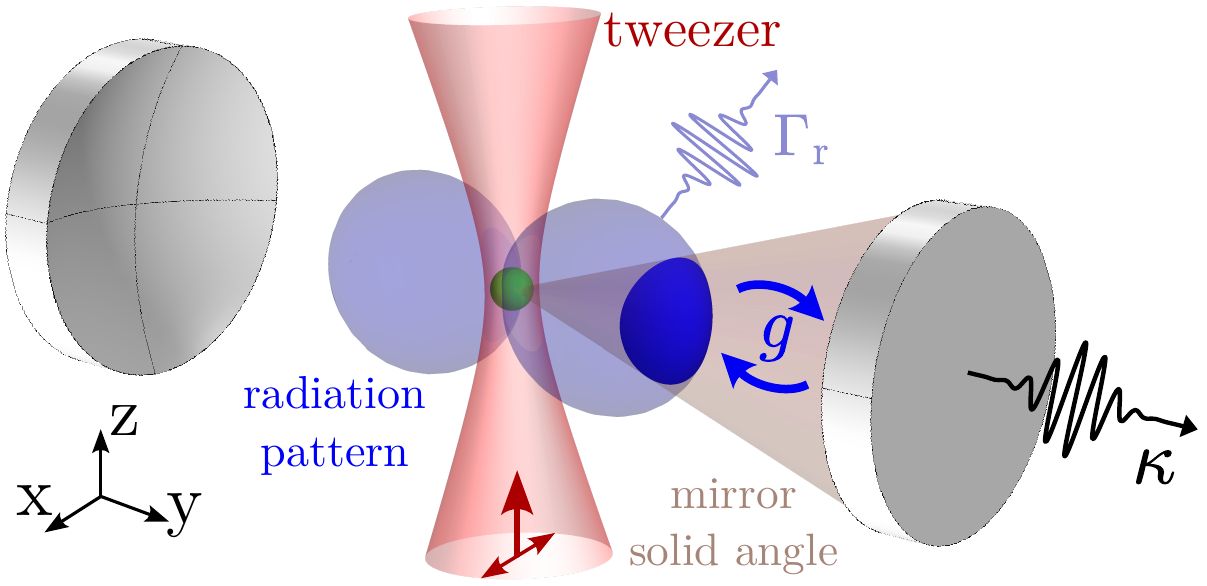}
    \caption{A sub-wavelength nanoparticle (green) is trapped by an optical tweezer (red).   
    In free space, the angular distribution of Stokes and anti-Stokes photons scattered by the center-of-mass degree of freedom follows the characteristic information radiation pattern~\cite{Tebbenjohanns_2019, Militaru_2022} (shown in blue for the motion along $y$). Inside a cavity, the recoil heating rate changes as (i) part of the scattered photons populate the cavity mode instead of free-space, and (ii) the scattering pattern itself is modified by Purcell effect.}
    \label{fig:fig1}
\end{figure}

Nanoparticles trapped in optical tweezers are a promising platform to explore macroscopic quantum physics~\cite{Millen_2020,CGB_review_2021,Stickler_2021}, as their center-of-mass motional degrees of freedom can be controlled at the quantum regime~\cite{Delic_2020, Magrini_2021, Tebbenjohanns_2021,Kamba_2022,Militaru_2022,Magrini_2022,Piotrowski_2023, Rossi_2025,  Troyer_2026, Kamba_2025, Skrabulis_2026} and, theoretically, also prepared in non-classical states such a large quantum superpositions~\cite{Romero_Isart_2010,Chang_2010,Romero_Isart_2011_prl,Romero_Isart_2011_pra,Roda_Llordes_2024,Neumeier_2024,Bemani_2025}. A promising route toward these goals consists in trapping the particle inside an undriven optical cavity, which becomes populated only by the tweezer light scattered off the nanoparticle~(\autoref{fig:fig1}). This so-called coherent-scattering configuration~\cite{Delic_2019, Windey_2019} results in strong optomechanical interaction between the nanoparticle and cavity mode, and enables efficient motional detection through the cavity output. Coherent scattering experiments enabled the first realization of ground-state motional cooling of levitated nanoparticles~\cite{Delic_2020}, later extended to multiple motional modes~\cite{Piotrowski_2023, Troyer_2026} and to librational degrees of freedom~\cite{Dania_2025, Troyer_2026}.

The possibility to prepare non-classical states  in any optical levitation platform is strongly limited by motional decoherence induced by laser shot noise, i.e., recoil heating. For a nanoparticle trapped in free space, recoil heating stems from  scattering of photons from the trapping tweezers into undetected free-space electromagnetic (EM) modes~\cite{Jain_2016}, and is well understood theoretically. Specifically, the recoil heating rate is computed perturbatively as a sum of scattering amplitudes over all the free-space EM modes~\cite{CGB_2019,Maurer_2023}. This approach cannot be extended to particles trapped inside a cavity, for two reasons: (i) computing all Maxwell eigenmodes in complex geometries is challenging, even for advanced numerical methods; (ii) the particle motion is usually strongly coupled to the narrow cavity mode, preventing a perturbative calculation. 
With quantitative predictions unavailable, current works approximate the recoil heating rate by its free-space value~\cite{Jain_2016, CGB_2019, Rudolph_2020,Rudolph_2021,Schafer_2021,Vijayan_2024,Troyer_2026}, ignoring its expected modification due to the modified local density of states (Purcell effect)~\cite{Novotny_2012}.
While this is expected to be approximately true for optical cavities whose cavity mirrors cover small solid angles, to assess and minimize decoherence in future experiments, the ability to quantitatively predict realistic recoil heating rates is needed.

In this Letter, we provide a general solution to quantify recoil heating rates in Fabry-Pérot cavities with arbitrary geometries and show that recoil heating decoherence can be suppressed by photonic structure engineering. To achieve this, we use the formalism of macroscopic quantum electrodynamics~\cite{Scheel_2008,buhmann2012a,buhmann2012b} to
derive a coherent scattering master equation for cavity and motional modes. The rates in such equations are functions of the electromagnetic Green's tensor and can thus be computed with standard EM solvers.
A crucial advance of our approach is the formal separation of the mode spectral density into a single resonance -- identified as the cavity mode and retained in the Hamiltonian -- and a background that can then safely be perturbatively eliminated. We achieve this through the few-mode quantization approach developed in quantum nanophotonics~\cite{Medina_2021, Sanchez-Barquilla_2022, Lednev_2024, Lednev_2025}. We show that in realistic, high-finesse microcavities, the recoil heating rate can be significantly modified with respect to its free-space value, providing a path toward decoherence engineering in levitated optomechanics. 

We consider a spherical nanoparticle with radius $R$, mass $M$, and polarizability $\alpha$, trapped by an optical tweezer with frequency $\omega_0=ck_0=2\pi c/\lambda_0$ which propagates along the $z$ axis and is polarized along $x$-axis (\autoref{fig:fig1}). The intensity maximum of the tweezer defines the origin of coordinates. The particle is surrounded by an arbitrary EM structure, characterized by the scalar relative permittivity and permeability profiles $\varepsilon(\mathbf{r},\omega)$ and $\mu(\mathbf{r},\omega)$. Within the point-dipole approximation for the nanoparticle, $k_0R \ll 1$, the system Hamiltonian reads~\cite{CGB_2019}
\begin{align}
    \hat{H} &=  \frac{\hat{\mathbf{P}}^2}{2M} +\hbar\sum_{\sigma=e,m} \int \mathrm{d}^3\mathbf{r} \int_0^\infty \mathrm{d}\omega\  \omega \hat{\mathbf{f}}_\sigma^\dagger(\mathbf{r},\omega) \cdot \hat{\mathbf{f}}_\sigma(\mathbf{r}, \omega)\nonumber\\&- \frac{\alpha}{2} (\mathbf{E}_{\mathrm{tw}}(\hat{\mathbf{R}},t)+\hat{\mathbf{E}}(\hat{\mathbf{R}}))^2,
    \label{eq:H_total}
\end{align}
where $\hat{\mathbf{P}}$ and $\hat{\mathbf{R}}$ are the nanoparticle's center-of-mass momentum and position operators.
The second term is the Hamiltonian of the EM field within the macroscopic quantum electrodynamics (QED) formalism. It is described by  the  ladder operators $\hat{\mathbf{f}}_\sigma(\mathbf{r}, \omega)$ and $\hat{\mathbf{f}}_\sigma^\dagger(\mathbf{r},\omega)$,  where $\sigma=e,m$ denotes electric and magnetic contributions. These operators are associated to the medium-assisted macroscopic EM modes, that is, the full eigenmodes of the coupled system formed by the EM field and the matter excitations in the surrounding materials. They obey bosonic commutation relations, $[\hat{\mathbf{f}}_{\sigma}(\mathbf{r},\omega),\hat{\mathbf{f}}_{\sigma'}(\mathbf{r}',\omega')]=\mathbb{0}$ and $[\hat{\mathbf{f}}_{\sigma}(\mathbf{r},\omega),\hat{\mathbf{f}}^\dagger_{\sigma'}(\mathbf{r}',\omega')]=\delta_{\sigma\sigma'} \delta(\omega-\omega') \delta(\mathbf{r} - \mathbf{r}')\mathbb{1}$, with $\mathbb{0}$ and $\mathbb{1}$ the $3\times 3$ zero and identity matrices, respectively. 

The third term in \autoref{eq:H_total} represents the optomechanical interaction with two electric fields. First, the classical field of the optical tweezer $\mathbf{E}_{\mathrm{tw}}(\mathbf{r},t)=\text{Re}[\mathcal{E}_{tw}(\mathbf{r})\mathbf{e}_xe^{-i\omega_0t}]$, with $\mathbf{e}_x$ being the unit vector along the x-axis and $\mathcal{E}_{tw}(\mathbf{r})$ 
an arbitrary mode profile. 
Its expression for a zeroth-order Gaussian beam -- corresponding to the case study below -- is given in the End Matter.
Second, the medium-assisted electric field, 
\begin{equation}\label{Eoperator}
    \hat{\mathbf{E}}(\mathbf{r}) = \!\sum_{\sigma=e,m} \!\int \mathrm{d}^3r' \!\int_0^\infty \!\!\mathrm{d}\omega\ \boldsymbol{\mathcal{G}}_\sigma(\mathbf{r},\mathbf{r}',\omega)\cdot\hat{\mathbf{f}}_\sigma(\mathbf{r}', \omega)  + \mathrm{H.c.}
\end{equation}
with the dyadic Green's tensors
$\boldsymbol{\mathcal{G}}_e(\mathbf{r},\mathbf{r}',\omega) = i(\omega/c)^2\allowbreak\sqrt{\hbar \text{Im}[\varepsilon(\mathbf{r}',\omega)]/(\pi\varepsilon_0)} \boldsymbol{\mathcal{G}}(\mathbf{r},\mathbf{r}',\omega)$
and $\boldsymbol{\mathcal{G}}_m^T(\mathbf{r},\mathbf{r}',\omega) = (\omega/c)\allowbreak\sqrt{\hbar \text{Im}[\mu^{-1}(\mathbf{r}',\omega)]/(\pi\varepsilon_0)} \nabla'\times\boldsymbol{\mathcal{G}}(\mathbf{r},\mathbf{r}',\omega)$, $\varepsilon_0$ the vacuum permittivity. The tensor $\boldsymbol{\mathcal{G}}(\mathbf{r},\mathbf{r}',\omega)$ is the solution of the vector Helmholtz equation, $[\nabla\times(\mu^{-1}(\mathbf{r},\omega)\nabla\times)\allowbreak-(\omega^2\varepsilon(\mathbf{r},\omega)/c^2)]\boldsymbol{\mathcal{G}}(\mathbf{r},\mathbf{r}',\omega)=\delta(\mathbf{r}-\mathbf{r}')\mathbb{1}$.
A core advantage of the macroscopic QED formalism is that all the information about the EM structure is encoded in the tensor $\boldsymbol{\mathcal{G}}(\mathbf{r},\mathbf{r}',\omega)$, which is accessible in standard EM solvers. 
The medium-assisted field \autoref{Eoperator} is the most general expression of the field in the presence of EM structures. In the conventional theoretical approach to coherent scattering, this field is phenomenologically split into a cavity- and a free-space contribution~\cite{CGB_2019}. We simplify the interaction term in the Hamiltonian (\ref{eq:H_total}) via the following standard procedure, see Ref.~\cite{CGB_2019} for details: 
(i) We expand all the field operators to first order in position operators $\hat{\mathbf{R}}$ around the coordinate origin (Lamb-Dicke approximation). This is valid provided that $\vert\mathbf{R}_0\vert,\langle \hat{\mathbf{R}}^2\rangle^{1/2} \ll \lambda_0$, where $\mathbf{R}_0$ is the nanoparticle's equilibrium position defined below, and assuming that only modes with frequencies near $\omega_0$ contribute appreciably to the optomechanical interaction. (ii) We neglect all terms stemming from the contribution $-(\alpha/2)\hat{\mathbf{E}}^2(\hat{\mathbf{R}})$, an approximation valid in the regime $\mathcal{E}_{tw}(0) \gg \vert \langle\hat{\mathbf{E}}(0)\rangle\vert$, i.e., for EM structures with not too narrow resonances~\cite{IurieInPreparation}. 
This approximation amounts to neglecting scattering of thermal photons off the nanoparticle -- always negligible within the point-dipole approximation for the nanoparticle~\cite{PflanzerPRA2012,Maurer_2023} -- as well as small structure-induced mechanical frequency shifts.
(iii) We perform a unitary transformation of the EM operators to a frame rotating at $\omega_0$, and neglect  all terms oscillating at $2\omega_0$ and $\omega_0+\omega$. This rotating wave approximation is valid for $\Omega_i,J_i(\omega\approx\omega_0)\ll\omega_0$, where $\Omega_i$ and $J_i(\omega)$ are the mechanical frequencies and the spectral densities for motion along axis $i=x,y,z$ defined below. (iv) We apply a displacement transformation $\hat{\mathbf{f}}_\sigma(\mathbf{r},\omega)\rightarrow \hat{\mathbf{f}}_\sigma(\mathbf{r},\omega)+\boldsymbol{\gamma}_\sigma(\mathbf{r,\omega})$ and $\hat{\mathbf{R}} \to \hat{\mathbf{R}} + \mathbf{R}_0$. The equilibrium values $\boldsymbol{\gamma}_\sigma(\mathbf{r,\omega})$ and $\mathbf{R}_0$ -- and, through Eq.~\eqref{Eoperator}, also the field amplitude $\langle\hat{\mathbf{E}}(0)\rangle$ -- are determined by imposing the zero-force condition, i.e. the absence of single-operator terms in the transformed Hamiltonian. The obtained expressions, given in the End Matter, can be used to confirm the validity of approximations (i-ii). 
 
The Hamiltonian can then be cast as $\hat{H}=\sum_{i=x,y,z}\hat{H}_i$, with
\begin{multline}\label{eq:H_final}
    \hat{H}_i/\hbar = \Omega_i\hat{b}_i^\dagger\hat{b}_i+\int_0^\infty d\omega (\omega-\omega_0)\hat{a}^\dagger_i(\omega)\hat{a}_i(\omega)
    \\
    +\int_0^\infty d\omega \sqrt{J_i(\omega)}\hat{q}_i\left(\hat{a}_i(\omega) + \hat{a}^\dagger_i(\omega)\right).
\end{multline}
Here, we have defined the motional ladder operators as $\hat{R}_i\equiv r_{i0}(\hat{b}_i+\hat{b}_i^\dagger)\equiv r_{i0} \hat{q}_i$ with $r_{i0}\equiv \sqrt{\hbar/(2M\Omega_i)}$ the zero-point motion along axis $i$. The expressions for $\Omega_i$ for a Gaussian tweezer profile can be found in Ref.~\cite{CGB_2019} and are given in the End matter. We have also defined the collective EM mode operators
\begin{multline}\label{interactingmodes}
    \hat{a}_i(\omega) \equiv \frac{-\alpha r_{i0}}{2\hbar \sqrt{J_i(\omega)}}
     \sum_{\sigma=e,m} \int \mathrm{d}^3r'\\ \frac{\partial}{\partial r_{i}}\left[\mathcal{E}_{tw}(\mathbf{r})\mathbf{e}_x\cdot\boldsymbol{\mathcal{G}}_\sigma(\mathbf{r},\mathbf{r}',\omega)\right]\Big|_{\mathbf{r}=0}\cdot\hat{\mathbf{f}}_\sigma(\mathbf{r}', \omega),
\end{multline}
known in nanophotonics as \textit{bright} or \textit{emitter-centered modes}~\cite{buhmann2012b,Hummer_2013,Rousseaux_2016,Dzsotjan_2016,Castellini_2018,Varguet_2019,feist2021}, which generalize the concept of \textit{interacting modes} introduced in free-space levitated optomechanics~\cite{Magrini_2021, Militaru_2022} and exactly recover them in the free-space limit. Hereafter, we assume for simplicity that the surrounding
structure shows inversion symmetry with respect to
all Cartesian planes, which results in these modes being orthogonal, i.e., $[\hat{a}_i(\omega),\hat{a}^\dagger_j(\omega')]=\delta_{ij}\delta(\omega-\omega')$~\footnote{For a non-inversion-symmetric structure, these modes are not orthogonal to each other and commute as  $[\hat{a}_i(\omega),\hat{a}^\dagger_j(\omega')]=\delta(\omega-\omega')J_{ij}(\omega)/\sqrt{J_i(\omega)J_j(\omega)}$~\cite{feist2021}, with $J_{ij}(\omega)$ defined as in \autoref{eq:J} but with the substitution $r_{i0}^2(\partial^2/\partial r_i\partial r_i') \to r_{i0}r_{j0}(\partial^2/\partial r_i\partial r_j')$}. In this case, the Hamiltonian (\ref{eq:H_final}) decouples into three independent optomechanical Hamiltonians for each motional mode $x,y,z$. For each of these modes, the optomechanical interaction is fully characterized by the spectral density,
    \begin{multline}
        J_{i}(\omega) \equiv\frac{\alpha^2r^2_{i0}\omega^2}{4\hbar\pi\varepsilon_0c^2}\frac{\partial^2}{\partial r_{i} \partial r_{i}'}\\  \left[\ \mathcal{E}_{tw}(\mathbf{r}) \mathrm{Im}\mathcal{G}_{xx}(\mathbf{r},\mathbf{r}',\omega) \mathcal{E}^*_{tw}(\mathbf{r}')\right]\Big|_{\mathbf{r}=\mathbf{r}'=\mathbf{R}_0}.
        \label{eq:J}
    \end{multline}
The above reformulation of the coherent scattering Hamiltonian is especially suited for numerical calculations, as we discuss in the following.

\begin{figure}[tb]
    \centering
    \includegraphics[width=0.95\linewidth]{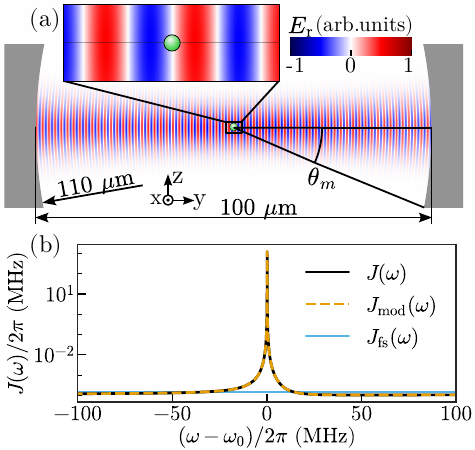}
    \caption{(a) Cavity geometry and radial electric field profile at the cavity resonance frequency (nanoparticle not to scale). The maximal field gradient along the cavity axis provides strong optomechanical coupling to the $y$-motion. For all figures we choose the tweezer and nanoparticle parameters of Ref. ~\cite{CGB_2019}. (b) Spectral density of the $y-$motion (black) and its two-mode fit $J_{\mathrm{mod}}$ (yellow) for $\theta_m=23.8^\circ$. The free-space spectral density $J_{\mathrm{fs}}$ is shown in blue.}
    \label{fig:fig2}
\end{figure}

As a case study, we apply our formalism to the configuration shown in ~\autoref{fig:fig2}(a). The tweezer spot lies at the center of a symmetric near-confocal Fabry–Pérot microcavity formed by identical metallic mirrors (modeled with constant permittivity $\varepsilon_m=-10^4$ for simplicity) and permeability $\mu_m=1$, curvature radius $R=110~\mathrm{\mu m}$, and separated by a 
length $L_c=100~\mathrm{\mu m}$. 
We assume rotational symmetry around the cavity axis ($y$), so that the angular size of the mirrors is quantified by the single angle $\theta_m$. 
We focus on the center-of-mass motion along the $y$ axis~\cite{CGB_2019,Delic_2020}, and thus drop hereafter the index $i$ and the summation over spatial directions in \autoref{eq:H_final}. 
The tweezer optical frequency is set at $\omega_0/2\pi=192.57$~THz ($\lambda_0=1556.8$~nm) such that the spectral density is peaked near the frequency of the anti-Stokes photons scattered by the nanoparticle, the usual configuration for cooling experiments~\cite{CGB_2019,Delic_2020}.
We use COMSOL Multiphysics~\cite{comsol} to compute the electric field generated by a point source in this geometry, and thus the Green's tensor $\mathcal{G}(\mathbf{r},\mathbf{r}';\omega)$~\cite{Novotny_2012}. From it, we evaluate the spectral density $J(\omega)$ via \autoref{eq:J} and display it in \autoref{fig:fig2}(b) (black curve) for $\theta_m=23.8^\circ$. The spectral density exhibits a narrow resonance, typically identified with a cavity mode. This is confirmed by the transverse field distribution at the peak frequency (\autoref{fig:fig2}(a)), which shows the characteristic Gaussian mode profile. The flat background of the spectral density in \autoref{fig:fig2}(b) describes coupling to a broad fluctuating continuum, giving rise to the recoil heating decoherence. Note that this background is not equal to its free-space value $J_\mathrm{fs}(\omega)$ (blue curve), already indicating a cavity-induced recoil heating modification.

The next key step to develop a coherent scattering master equation is to recast the continuum Hamiltonian~(\ref{eq:H_final}) in the conventional optomechanical form, namely a two-mode master equation for the mechanical and cavity modes where the EM continuum is accounted for perturbatively~\cite{CGB_2019,Rudolph_2020,Toros_2020,Toros_2021,Rudolph_2021,Brandao_2021}. Adiabatic elimination of all continuum modes~\cite{Dung_2002,Seberson_2020,Cerchiari_2021,JakubecPRA2025,Gajewski_2025} is not possible for this system, as the spectral density cannot be perturbatively treated near the cavity resonance.  We thus make use of the few-mode quantization approach~\cite{Medina_2021, Sanchez-Barquilla_2022, Lednev_2024, Lednev_2025}, which provides an exact mapping of the mode continuum $\hat{a}(\omega)$ onto a set of discrete lossy modes $\hat{\mathbf{c}}=\{\hat{c}_1,...,\hat{c}_N\}$ coupled to each other, i.e., a mapping to the rotating-frame master equation
\begin{align}
    \dot{\hat{\rho}} = -i\hbar^{-1}\left[\hat{H}_\mathrm{mod},\hat{\rho}\right] + \sum_{\beta=1}^N \kappa_\beta \mathcal{L}_{\hat{c}_\beta,\hat{c}_\beta}[\hat{\rho}],
    \label{eq:rhomod}
\end{align}
where $\mathcal{L}_{\hat{O}_i,\hat{O}_j}[\hat{\rho}]\equiv\hat{O}_i\hat{\rho}\hat{O}^\dagger_j-(1/2)\{\hat{O}^\dagger_j\hat{O}_i,\hat{\rho}\}$ and $\hat{H}_\mathrm{mod}/\hbar=\allowbreak\hat{\mathbf{c}}^\dagger\cdot\Lambda\cdot\hat{\mathbf{c}}+ \Omega_y\hat{b}^\dagger\hat{b} +\sum_{\beta} g_\beta(\hat{c}_\beta + \hat{c}_\beta^\dagger)\hat{q}$, with $\Lambda_{\beta\beta'}=\omega_{\beta\beta'}-\omega_0\delta_{\beta\beta'}$. The rates $\omega_{\beta\beta},\kappa_\beta$, and $g_\beta$ are the frequency, decay rate, and optomechanical coupling rate of optical mode $\beta$ to the center-of-mass motion, whereas intermode coupling is described by $\omega_{\beta\beta'}$ ($\beta\ne\beta'$). 
This discrete model is formally equivalent to a continuum model with spectral density $J_{\mathrm{mod}}(\omega)=\pi^{-1}\mathbf{g}\cdot\mathrm{Im}[(\Lambda-i\text{diag}[\boldsymbol{\kappa}/2]-\omega \mathbb{1}_{N\times N})^{-1}]\cdot\mathbf{g}^T$. Hence, to ensure its equivalence to the original Hamiltonian \autoref{eq:H_final}, we fit the function $J_{\mathrm{mod}}(\omega)$ to the spectral density $J(\omega)$ (\autoref{eq:J}) over the frequency range of interest, using as fitting parameters the rates $\{g_\alpha$, $\omega_{\alpha\beta}$, $\kappa_\alpha\}$. 
The accuracy of this fit can be arbitrarily increased by increasing the number of fitting modes $N$.

The spectral density of the system under consideration [\autoref{fig:fig2}(b)] can be accurately fit using only two modes $\{\hat{c}_1,\hat{c}_2\}$ (see yellow curve). The first mode has the frequency of the spectral density peak and narrow linewidth, whereas the second mode is broad, i.e. $\kappa_2 \gg \kappa_1, g_{1},g_2, \vert\omega_2-\omega_0\vert$, and represents the whole continuum of available propagating modes. This separation of energy scales allows for a standard adiabatic elimination of the second mode under a Born-Markov approximation~\cite{Brion_2007, CGB_2024_tutorial, Ben-Asher_2026}. This procedure results in the final coherent scattering master equation for the $y-$motional mode $\hat{b}$ and the cavity resonance $\hat{c}_1$ (hereafter labelled $\hat{c}$ for simplicity),
\begin{equation} \label{eq:rho_sm}
    \dot{\hat{\rho}}=-\frac{i}{\hbar}[\hat{H}_{\rm CS},\hat\rho]+\kappa\mathcal{L}_{\hat{c},\hat{c}}[\hat\rho]-\frac{\Gamma}{2}\left[\hat{q},\left[\hat{q},\hat\rho\right]\right],
\end{equation}
with a Hamiltonian $\hat{H}_{\rm CS}/\hbar=(\omega_c-\omega_0)\hat{c}^\dagger\hat{c}+\Omega_y\hat{b}^\dagger\hat{b}+g(\hat{c}+\hat{c}^\dagger)\hat{q}$ and where we identify $\omega_c=\omega_1$, $g=g_1$, the renormalized cavity dissipation rate $\kappa\equiv \kappa_1+4\omega_{12}^2/\kappa_2$, and the recoil heating rate $\Gamma \equiv 4 g_2^2/\kappa_2$. In \autoref{eq:rho_sm}, we have neglected dissipative motion-cavity coupling terms as the associated rates are much smaller than the coherent coupling $g$~\footnote{These terms arise from the finite overlap between the field modes responsible for cavity loss and for Stokes/antiStokes scattering, which in arbitrary structures can be very large. Our model fully captures this effect, as opposed to previous models that neglect it by necessity~\cite{CGB_2019}.}. \autoref{eq:rho_sm} is the core result of this work: it has the usual coherent scattering form~\cite{CGB_2019} but, crucially, all the rates are exactly computable from the spectral density. 

\begin{figure}[tb]
    \centering
    \includegraphics[width=0.95\linewidth]{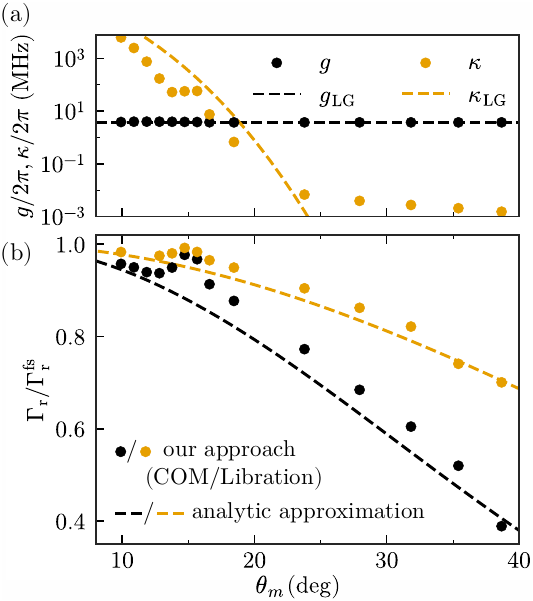}
    \caption{(a) Numerically obtained optomechanical coupling rate $g$ and cavity linewidth $\kappa$ versus mirror angular size $\theta_m$. The dashed lines show independent analytical calculations which describe the cavity through a single Laguerre-Gaussian mode, see text for details. (b) Recoil heating decoherence normalized to its free-space value versus mirror solid angle, for center-of-mass motion and for libration (black and orange dots). The dashed lines indicate the geometric approximation \autoref{rhapprox} for each case. The peak at $\theta_m\approx15^\circ$ 
    stems from additional, higher-order, broad resonances that are supported by the cavity at these angles and that modify the density of states (see End Matter for details).}
    \label{fig:fig3}
\end{figure}

To further characterize our model, we first show the numerically obtained cavity decay and optomechanical coupling rates, $\kappa$  and $g$, in \autoref{fig:fig3}(a) as a function of angular mirror size $\theta_m$. 
The monotonic decrease of $\kappa$ at small $\theta_m$ is due to diffractional losses, and is well captured by the analytical expression for a zeroth-order Laguerre-Gaussian cavity mode  (orange dashed line), see End Matter. Similarly, the coupling rate g (black dashed line) quantitatively agrees with the expression derived in the literature for a zeroth-order Laguerre-Gaussian cavity mode, i.e., $g_{\rm LG}=(2\hbar\varepsilon_0 V_c/\omega_c)^{-1/2}\alpha E_0 \omega_c r_{y0}/(2c)$~\cite{CGB_2019} with $V_c=\pi W_c^2L_c/4$ and  the cavity mode waist $W_c$. This waist is extracted from the data of \autoref{fig:fig2}(a) by fitting a Gaussian function to the field distribution at plane $y=0$. The results in \autoref{fig:fig3}(a) confirm that our approach quantitatively and accurately describes  coherent scattering dynamics. We remark that the contribution of the eliminated mode $\hat{c}_2$ to both $g$ and $\kappa$ is negligible, confirming that the main role of this broadband mode is to induce recoil heating decoherence.

The numerically calculated recoil heating rate normalized to its free-space value, $\Gamma/\Gamma_{\rm fs}$, is shown in \autoref{fig:fig3}(b) (black dots). As mirror angular size increases, the EM modes become increasingly different from their free-space analogues, and the recoil heating is suppressed. 
The decreasing trend can be understood geometrically as the result of the mirrors directing an increasingly large fraction of the Stokes- and anti-Stokes-scattered photons into the cavity mode instead of into free space.
To confirm this intuition we compare our exact results with the expression
\begin{equation}\label{rhapprox}
    \Gamma_{\rm an} =  \Gamma_{\rm fs}\frac{\int_{\Omega_{\rm open}}\Phi_y(\Omega)d\Omega}  {\int_{4\pi}\Phi_y(\Omega)d\Omega},
\end{equation}
where $\Gamma_{\rm fs}$ is the free-space recoil heating, $\Omega$ represents a solid angle variable, and $\Phi_y(\Omega)=\sin^2\phi\ \sin^2 \theta (1 - \cos^2\phi\ \sin^2 \theta)$ is the so-called free-space information radiation pattern, i.e., the spatial distribution of the Stokes and anti-Stokes photons scattered by the $y-$motional mode~\cite{Tebbenjohanns_2019, Maurer_2023, Magrini_2018, Militaru_2022,CGB_2023,Hupfl_2024} (blue surface in \autoref{fig:fig1}). \autoref{rhapprox} thus describes the recoil heating decoherence caused only by photons scattering in directions that are not physically blocked by the mirrors (solid angle $\Omega_{\rm open}$), and is analytically calculated as $\Gamma_{\rm an} =  \Gamma_{\rm fs}\cos^3(\theta_m)\left[3\cos(2\theta_m)+13\right]/16$. As shown by the black dashed line in \autoref{fig:fig3}(b), this expression is a good approximation to the exact result, indicating that for this specific system this geometric effect is the main responsible for the decoherence suppression. 

A second effect causing a change in the recoil heating is the strong modification of the density of states, which manifests in the maximum at $\theta_m\approx 15^\circ$. As shown in the End Matter, at this angle the mirrors are large enough to sustain an additional, higher-order resonance.
At larger $\theta_m$, this resonance does not contribute to the dynamics as it is narrow and far detuned with respect to the main resonance at $\sim \omega_c$. However, at $\theta_m<20^\circ$ the mode is poorly confined and thus spectrally wide enough to provide an additional decoherence channel for the center-of-mass motion. Finally, to illustrate the generality of our model, we extend it beyond center-of-mass motion and compute the recoil heating rate of the librational mode of an anisotropic nanoparticle along $z$, shown in orange in \autoref{fig:fig3}(b) (see End Matter for details). Recoil heating decoherence is also suppressed, although this effect is less pronounced than for center-of-mass motion due to the more isotropic information scattering pattern~\cite{CGB_2023}. The possibility of computing  recoil heating rate exactly and to suppress it via nanostructure engineering -- an idea that extends to other decoherence mechanisms~\cite{AgreniusInPreparation} -- is core to coherent-scattering levitodynamics and the main result of this work.

In conclusion, we have developed a quantum theory of levitated optomechanics via coherent scattering, using the macroscopic QED and few-mode quantization formalisms. Our model reproduces known results and goes beyond current approaches, as it allows to quantitatively compute recoil heating decoherence for arbitrary geometries. This capability is essential for realizing macroscopic quantum states of optically levitated nanoparticles. Our general approach can be easily applied to EM structures beyond conventional Fabry–Pérot cavities~\cite{Neumeier_2015, Mestres_2016, Magrini_2018, Monterrosas-Romero_2023, Alavi_2025}, as the expressions for all dynamical rates are particularly suited for numerical evaluation using conventional Maxwell solvers. Combined with the straightforward extension to multiple particles and to arbitrary excitation beam configurations, our work provides the theoretical toolbox for ``coherent-scattering nanoengineering'' in levitodynamics, namely the nanophotonics-inspired design of structures that minimize recoil heating motional decoherence or maximize directionality of scattered photons for efficient motional detection. 

\begin{acknowledgments}
M.L. and J.F. acknowledge support by the Spanish Ministerio de Ciencia y Universidades — Agencia Estatal de Investigación through the FPI Grant PRE2021-098978, the grant PID2024‐161142NB‐I00, and the grant CEX2023‐001316‐M through the María de Maeztu program for Units of Excellence in R\&D. U.D. acknowledges the support of the Austrian Science Fund (FWF) 10.55776/STA175. C.G.B. was supported by the Austrian Science Fund (FWF) [10.55776/COE1 and 10.55776/PIN3404324]. 
\end{acknowledgments}
\bibliography{bib.bib}

\clearpage
\section{End Matter}

\subsection{Tweezer with a Gaussian profile}

In accordance with most experiments, for the case study shown in Figs.~\ref{fig:fig2}-\ref{fig:fig3} we assume that the tweezer mode profile is given by the free-space zeroth-order Laguerre-Gaussian beam, i.e, we use $\mathcal{E}_{tw}(\mathbf{r})= E_0(W_t/W(z))\allowbreak \exp[i(k_0z-\phi_t(\mathbf{r}))-(x^2+y^2)/W^2(z)]$~\cite{Novotny_2012, CGB_2019}, where $E_0$ is the field amplitude, $W(z)=W_t\sqrt{1+(z/z_R)^2}$ with $W_t$ the tweezer waist, and $z_R=\pi W_t^2/\lambda_0$  the Rayleigh range, and with a Gouy phase $\phi_t(\mathbf{r})=\arctan(z/z_R)-(k_0 z/2)(x^2+y^2)/(z^2+z_R^2)$. We also assume that the nearby EM structures do not modify significantly this free-space shape~\footnote{Modifications, however, can easily be included by substituting the Gaussian field profile by the field profile in the presence of the structure, which can be obtained by solving the corresponding classical scattering problem.}. For this mode profile, the center-of-mass mechanical frequencies read $\Omega_x=\Omega_y=\Omega_z\pi\sqrt{2}W_t/\lambda_0=\sqrt{\alpha E_0^2/(MW_t^2)}$.

\subsection{Equilibrium amplitudes of the operators}

As mentioned in the main text, in order to simplify the Hamiltonian we applied a displacement transformation $\hat{\mathbf{f}}_\sigma(\mathbf{r},\omega)\rightarrow \hat{\mathbf{f}}_\sigma(\mathbf{r},\omega)+\boldsymbol{\gamma}_\sigma(\mathbf{r,\omega})$ and $\hat{\mathbf{R}} \to \hat{\mathbf{R}} + \mathbf{R}_0$. Inserting these expressions into Hamiltonian and imposing all the Hamiltonian terms with single operators $\hat{\mathbf{f}}_\sigma(\mathbf{r},\omega)$ and $\hat{\mathbf{R}}$ to turn to 0, we obtain a coupled system of equations for $\mathbf{R}_0$ and $\boldsymbol{\gamma}_\sigma(\mathbf{r,\omega})$. Their solution gives us the following equilibrium position of the nanoparticle and coherent amplitude of the electric field (to get the latter, one needs to insert the found value of $\boldsymbol{\gamma}_\sigma(\mathbf{r,\omega})$ into \autoref{Eoperator}):
\begin{subequations}
    \begin{equation}
        \mathbf{R}_0=\frac{\mathbf{e}_z\partial_{z'}O_{xx}}{(E_0\lambda_0/\pi W_t^2)^2-2\partial_z\partial_{z'}O_{xx}\vert_{\mathbf{r}=\mathbf{r}'=0}}
    \end{equation}
    \begin{equation}
        \langle\hat{\mathbf{E}}(0)\rangle =\mathbf{O}_x(0,0)\text{Re}\left[\left(1+2|\mathbf{R}_0|\frac{\partial}{\partial z}\right)\mathcal{E}^*_{tw}(\mathbf{r})\right]_{\mathbf{r}=0},
\end{equation}
\end{subequations}
where we define
\begin{equation*}
    \mathbf{O}(\mathbf{r},\mathbf{r'})=\frac{\alpha \mathcal{E}_{tw}(\mathbf{r}) \mathcal{E}^*_{tw}(\mathbf{r}')}{2\pi\varepsilon_0c^2}\mathcal{P}\int_0^\infty \frac{d\omega\ \omega^2}{\omega-\omega_0}\text{Im}\boldsymbol{\mathcal{G}}(\mathbf{r},\mathbf{r}',\omega)
\end{equation*}
with $\mathcal{P}$ indicating the Cauchy principal value. To derive these expressions, we have assumed that the surrounding structure shows inversion symmetry with respect to all Cartesian planes, but the generalization to arbitrary geometries is straightforward. 

\begin{figure}[htbp]
    \centering
    \includegraphics[width=\linewidth]{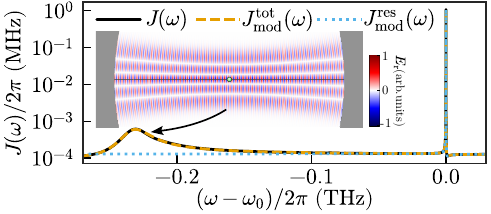}
    \caption{ Spectral density perceived by the $y$-motion (black) for $\theta_m=16.6^\circ$; and its fits in extended (yellow) and resonant (blue) regions, respectively. (inset) Spatial distribution of the radial component of the electric field $E_\mathrm{r}$ of the detuned cavity mode.}
    \label{fig:detuned_mode}
\end{figure}

\subsection{Estimation for cavity decay rate}

For the mirror permittivity chosen in this work ($\varepsilon=-10^4$), the mirrors are considered lossless and with negligible skin depth. Hence, for sufficiently small mirrors,  photon leakage -- and thus cavity decay rate -- will be dominated by diffraction at the mirrors edges. The fraction of energy lost to diffraction losses is
\begin{equation}\label{eq:loss}
    \mathrm{Loss}= \frac{I_{\mathrm{out}}}{I_{\mathrm{total}}} = \frac{\int_a^\infty |E(\rho;y_m)|^22\pi\rho d\rho}{\int_0^\infty|E(\rho;y_m)|^22\pi\rho d\rho} = e^{-2a^2/w_m^2},
\end{equation}
where $I_{out}$ and $I_{total}$ denote the intensities of the light leaking out of the cavity at the mirror edges and the intensity of the light arriving at the mirror, respectively; $E(\rho;y)$ is the electric field of the cavity mode, which we model as a Laguerre-Gaussian beam propagating along the $y$-axis, such that $|E(\rho;y)|^2 \propto \exp[-2\rho^2/W^2(y)]$, with $\rho=\sqrt{x^2+z^2}$ and $W(y)$ being the waist of the beam at the distance $y$ from the focus. $y_m$ is the distance between the mirror and the cavity center, $w_m=W(y_m)$, and $a$ is the mirror radius. The latter is connected with angular mirror size as $(L/2-R+\sqrt{R^2-a^2})\tan\theta_m=a$. Assuming that diffraction losses dominate, the fraction of the energy that remains in the cavity after a full round trip (i.e. a trip involving one reflection from each mirror) is
\begin{equation}
    P_s = (1-\mathrm{Loss})^2.
\end{equation}
Using this expression, we compute the cavity finesse~\cite{Steck_2021}
\begin{equation}
    \mathcal{F} = \frac{\pi P_s^{1/4}}{1-P_s^{1/2}}.
\end{equation}
The cavity finesse is by definition equal to the ratio of the cavity free spectral range to the resonance linewidth, $ \mathcal{F} = \pi c/(L_c\kappa)$. Combining these two expressions for $\mathcal{F}$ and introducing \autoref{eq:loss}, we obtain
\begin{equation}
    \kappa = \frac{ce^{-2a^2/w_m^2}}{L_c\sqrt{1-e^{-2a^2/w_m^2}}},
\end{equation}
which is the expression shown by the dashed orange curve in \autoref{fig:fig3}(a).

\subsection{Higher order modes}

To substantiate our understanding of the peak observed at \autoref{fig:fig3}(b), we performed additional simulations probing the spectral density on a wider range of frequencies around the cavity resonance. For $12^\circ\lesssim\theta_m\lesssim20^\circ$ it features an additional wide peak (See example in \autoref{fig:detuned_mode}). This corresponds to the emergence of a higher-order Gaussian mode, as evidenced by the spatial distribution of the radial field at the central frequency of the wide peak [See inset in \autoref{fig:detuned_mode}].
Higher-order transverse modes are known to arise in Fabry–Pérot cavities~\cite{Steck_2021, Novotny_2012} and, due to their larger spatial extent, they exhibit larger diffractional losses than the fundamental mode. This means that they form only in cavities with large enough mirrors. For the mirror sizes comparable to their spatial width, these modes appear with large losses, as observed in our simulations for mirror sizes $\theta_m\approx15^\circ$. Due to these big losses, the spectral tails of these modes contribute to the spectral density background near the fundamental cavity resonance and thus affect the recoil heating decoherence rate. As the mirror size $\theta_m$ increases, these modes become spectrally narrower, making their contribution at the fundamental cavity resonance negligible.

The appearance of the second peak in \autoref{fig:detuned_mode} implies that to reproduce well the spectral density, one needs to introduce more modes into the few-mode fit. Specifically, instead of the 2 modes needed to reproduce all the features near the cavity resonance (blue line in \autoref{fig:detuned_mode}), to fit the spectral density in the extended frequency range four modes are needed (yellow). One of them is broad, and, as discussed in the main text, it can be adiabatically eliminated. The other two are significantly detuned from the laser frequency and consequently can be also adiabatically eliminated using a Schrieffer-Wolff transformation~\cite{CGB_2024_tutorial}. As a result, one obtains the same master equation \autoref{eq:rho_sm} with parameters which negligibly differ from the ones obtained in the main text. The total recoil heating rate in this scenario will be a sum of contributions from elimination of broad mode and detuned modes. The former one dominates, although the latter gives a notable contribution of few percent, confirming the fact that the peak at \autoref{fig:fig3}(b) is caused by detuned cavity modes. In the end, we note that the recoil heating rate obtained from fitting only narrow frequency range of $J(\omega)$ (see blue line in \autoref{fig:detuned_mode}) is the same as the total one obtained from extended frequency range (orange line).

\subsection{Interaction between librational degrees of freedom and EM environment}

Our formalism has the same form when applied to librations of an anisotropic particle~\cite{Schafer_2021, Rusconi_2022, CGB_2023}. We consider an anisotropic nanoparticle trapped by an optical tweezer polarized along $x$-axis and propagating along $z$, embedded into arbitrary EM environment. The nanoparticle is characterized by its mass $M$ and polarizability tensor which in the body frame reads
\begin{equation}
    \boldsymbol{\alpha}_{\mathrm{bf}} =
    \begin{pmatrix}
        \alpha_1& 0 & 0\\
        0 & \alpha_2 & 0\\
        0 & 0 & \alpha_3
    \end{pmatrix},
\end{equation}
with $\alpha_3>\alpha_1=\alpha_2$. Dipole forces align the nanoparticle's long axis along the tweezer polarization. If one neglects the center-of-mass dynamics, the motion of the nanoparticle is restricted to small angular displacements (librations) $\eta_y,\eta_z\ll1$, corresponding to rotations along the $y$ and $z$ axes, respectively. The Hamiltonian of such a system reads
\begin{align}
&\hat{H} = \hat{H}_f  + \frac{\hat{\mathbf{L}}^2}{2I}\nonumber\\
&-(\hat{\mathbf{E}}(\mathbf{R}_0) + \mathbf{E}_{tw}(\mathbf{R}_0))\frac{\boldsymbol{\alpha}_{\mathrm{lf}}}{2}(\hat{\mathbf{E}}(\mathbf{R}_0) + \mathbf{E}_{\mathrm{tw}}(\mathbf{R}_0)),
\label{eq:H_N_libr}
\end{align}
where $\hat{\mathbf{L}}$ and $I$ are the angular momentum operator and the moment of inertia of the nanoparticle. $\boldsymbol{\alpha}_{\mathrm{lf}}$ is the polarizability in the laboratory frame. Its connection with $\boldsymbol{\alpha}_{\mathrm{bf}}$ in the limit of $\eta_y,\eta_z\ll1$ is $\boldsymbol{\alpha}_{\mathrm{lf}}\approx\boldsymbol{\alpha}_{\mathrm{bf}}+(\alpha_3-\alpha_1)\mathbf{M}(\eta_y,\eta_z)$ with
\begin{equation}
    \mathbf{M} =
    \begin{pmatrix}
        -(\eta_z^2+\eta_y^2)& \eta_y & \eta_z\\
        \eta_y & \eta_y^2 & \eta_y\eta_z\\
        \eta_z & \eta_y\eta_z & \eta_z^2
    \end{pmatrix}
\end{equation}
being the rotation matrix. The term proportional to $\boldsymbol{\alpha}_{\mathrm{bf}}$ does not affect the rotational dynamics, so it can be dropped from the Hamiltonian. The remaining term can be simplified using the same standard steps (i)-(iv) discussed in the main text. This leads to a Hamiltonian of the same form as \autoref{eq:H_final}
\begin{multline}\label{eq:H_final_rot}
    \hat{H}/\hbar = \sum_{i=y,z}\Omega_i\hat{b}_i^\dagger\hat{b}_i+\int_0^\infty d\omega (\omega-\omega_0)\hat{a}^\dagger_i(\omega)\hat{a}_i(\omega)
    \\
    +\int_0^\infty d\omega \sqrt{J_i(\omega)}\hat{q}_i\left(\hat{a}_i(\omega) + \hat{a}^\dagger_i(\omega)\right),
\end{multline}
where  $\Omega=\sqrt{\Delta\alpha/(2I)}\left|E_0\right|$ is the librational frequency, $\Delta\alpha=\alpha_3-\alpha_1$ and $\hat{b}_i$ and $b_i^\dagger$ are the ladder operators for the phonons which characterize the libration along $i$-axis.
The interaction is characterized by the librational spectral density
\begin{equation}
    J_{i}(\omega) = \frac{\Delta\alpha^2\xi_{0}^2\omega^2|E_0|^2}{4\hbar\pi\varepsilon_0c^2} \mathrm{Im}\mathcal{G}_{ii}(\mathbf{R}_0,\mathbf{R}_0,\omega),
\end{equation}
and the generalized interacting modes read
\begin{align}
    &\hat{a}_i(\omega) = - \frac{\Delta\alpha \xi_{0}E_0}{2\hbar \sqrt{J_i(\omega)}}\nonumber\\
    &\sum_{\lambda=e,m} \int \mathrm{d}^3r' \left[\mathbf{e}_i\cdot\boldsymbol{\mathcal{G}}_\lambda(\mathbf{R}_0,\mathbf{r}',\omega)\right]\cdot\hat{\mathbf{f}}_\lambda(\mathbf{r}', \omega),
\end{align}
where $\xi_{0}=\sqrt{\hbar/(2I\Omega)}$ is zero-point angular displacement. 

Since the structure of this Hamiltonian is identical to that of the center-of-mass motion, the subsequent steps can be implemented in the same way to obtain the recoil heating decoherence rate. 
In this case, the geometric approximation \autoref{rhapprox} can be also computed analytically using the corresponding information radiation pattern, $\Phi^{l}_z = \sin^2\theta$~\cite{CGB_2023}, and reads
\begin{equation}
    \Gamma^{\mathrm{an},l}_{\mathrm{r}} = \Gamma^\mathrm{fs}_{\mathrm{r}}\left[15\cos(\theta_m)+\cos(3\theta_m)\right]/16.
\end{equation}
This expression is shown by the dashed orange curve in \autoref{fig:fig3}(b).

\end{document}